\documentclass[12pt]{article}
\usepackage{psfig}
\begin{document}

\def\lesssim{\mathrel{\mathpalette\vereq<}}
\def\gtrsim{\mathrel{\mathpalette\vereq>}}
\makeatletter
\def\vereq#1#2{\lower3pt\vbox{\baselineskip1.5pt \lineskip1.5pt
\ialign{$\m@th#1\hfill##\hfil$\crcr#2\crcr\sim\crcr}}}
\makeatother

\newcommand{\rem}[1]{{\bf #1}}

\renewcommand{\thefootnote}{\fnsymbol{footnote}}
\setcounter{footnote}{0}
\begin{titlepage}
\begin{center}

\hfill    UCB-PTH-98/54\\
\hfill    LBNL-42512\\
\hfill    hep-ph/9811299\\
\hfill    November 9, 1998\\

\vskip .5in

{\Large \bf U(2) and Maximal Mixing of $\nu_{\mu}$
\footnote
{This work was supported in part by the U.S. 
Department of Energy under Contracts DE-AC03-76SF00098, in part by 
the 
National Science Foundation under grant PHY-95-14797.}
}

\vskip .50in

Lawrence J. Hall and Neal Weiner

\vskip 0.05in

{\em Department of Physics\\
     University of California, Berkeley, California 94720}

\vskip 0.05in

and

\vskip 0.05in

{\em Theoretical Physics Group\\
     Ernest Orlando Lawrence Berkeley National Laboratory\\
     University of California, Berkeley, California 94720}

\vskip .5in

\end{center}

\vskip .5in

\begin{abstract}
A $U(2)$ flavor symmetry can successfully describe the 
charged fermion masses and mixings, and 
supress SUSY FCNC processes, making it a viable candidate for a theory
of flavor. We show that a direct application of this $U(2)$ flavor 
symmetry automatically predicts a mixing of $45^o$ for $\nu_\mu \Rightarrow 
\nu_s$, where $\nu_s$ is a light, right-handed state. The 
introduction 
of an additional flavor symmetry acting on the right-handed neutrinos makes 
the 
model phenomenologically viable, explaining the solar neutrino
deficit as well as the atmospheric neutrino anomaly, while giving a
potential hot dark matter candidate and
 retaining the theory's predictivity in the quark sector.

\end{abstract}
\end{titlepage}

\renewcommand{\thepage}{\arabic{page}}
\setcounter{page}{1}
\renewcommand{\thefootnote}{\arabic{footnote}}
\setcounter{footnote}{0}

\section{Introduction}

The pattern and origin of the quark and lepton masses and mixings
remains a challenging question for particle physics.
Although a detailed description of this pattern requires a theory of
flavor with a certain level of complexity, the gross features may be
described simply in terms of a flavor symmetry and its sequential 
breaking.

One simple flavor structure is motivated by four facts about flavor:
\begin{itemize}
\item The quarks and leptons fall into three generations,
$\psi_{1,2,3}$, each of which may eventually have a unified 
description.
\item The top quark is sufficiently heavy, that any flavor symmetry
which acts on it non-trivially must be strongly broken.
\item The masses of the two light generations imply a phenomenological
description in terms of small dimensionless parameters, $\{ \epsilon
\}$.
\item In supersymmetric theories, flavor-changing and $CP$
violating  phenomena suggest that the squarks and sleptons of the
first two generations are highly degnerate.
\end{itemize}
It is attractive to infer that, at least at a phenomenological
level, there is a non-Abelian flavor symmetry which divides the three
generations according to
\begin{equation}
 2 \oplus 1: \hspace {0.5in} \psi_a \oplus \psi_3, \,\,a = 1,2.
\label{eq:twoplusone}
\end{equation}
The four facts listed above follow immediately from such a structure,
with $\{ \epsilon \}$ identified as the small symmetry breaking
parameters of the non-Abelian group. These control both the small
values for quark masses and mixing angles, and also the small
fractional non-degeneracies of the scalars of the first two
generations.

The Super-Kamiokande collaboration has provided strong evidence for an
anomaly in the flux of atmospheric neutrinos, which may be interpreted
as large angle oscillations of $\nu_\mu$ predominantly either to
$\nu_\tau$ or to $\nu_s$, a singlet neutrino \cite{sk}. 
This observation provides a challenge to the non-Abelian $2 \oplus 1$
structure:
\begin{itemize}
\item $\nu_\tau$ is expected to have a very different mass from that 
of
$\nu_{e,\mu}$, and to only weakly mix with them.
\item If the atmospheric oscillation is $\nu_\mu \rightarrow \nu_s$, 
what is the identity 
of this new singlet state, why is it light, and how could it fit 
into 
the $2 \oplus 1$ structure?
\end{itemize} 

There are a variety of possible reactions to this challenge. One 
possibility is to drop the $2 \oplus 1$ idea; perhaps the $CP$ and flavor 
violating
problems of supersymmetry are solved by other means, or perhaps
supersymmetry is not relevant to the weak scale.
Another option is to retain the $2 \oplus 1$ structure for quarks, but not for
leptons, where the flavor changing constraints are much weaker.

In this paper we study theories based on the flavor group $U(2)$,
which immediately yields the structure (\ref{eq:twoplusone}), giving
the $2 \oplus 1 $ structure to both quarks and leptons \cite{PT}.
The masses and mixings of the charged fermions and scalars resulting
fron $U(2)$ have been
studied in detail, and significant successes have been identified 
\cite{u2}.
We add a right-handed neutrino to each generation, 
and find that the
symmetry structure of the neutrino mass matrix automatically chooses
$\nu_\mu$ to be a pseudo-Dirac state coupled to one of the
right-handed neutrinos, resulting in $\nu_\mu \rightarrow \nu_s$ with
a mixing angle close to $45^\circ$.

\section{U(2) Theories of Quark and Charged Lepton Masses.}

The most general $U(2)$ effective Lagrangian for charged fermion
masses, at leading order in the $U(2)$ breaking fields, is
\begin{equation}
\mathcal{L} = \psi_3 \psi_3 h + {1 \over M} \left( \
\psi_3 \phi^{a} \psi_a h + \psi_a (S^{ab} + A^{ab}) \psi_b h \right)
\label{eq:basicu2}
\end{equation}
where $\phi^a$ is a doublet, $S^{ab}$ a symmetric triplet, $A^{ab}$ an
antisymmetric singlet of $U(2)$, and $h$ are Higgs doublets.
Coupling constants have been omitted, and $M$ is a flavor physics 
mass scale.
An entire generation is represented by $\psi$, so that each operator 
contains terms in up, down and charged lepton
sectors, but unification is not assumed. For example, this theory 
follows 
from a renormalizable Froggatt-Nielsen model on integrating 
out 
a single heavy vector $U(2)$ doublet of mass $M$ (see the second of 
\cite{u2}).

The hierarchical pattern of masses and mixings 
for charged fermions is generated by breaking $U(2)$
first to $U(1)$ with vevs $\phi^2, S^{22} \approx \epsilon M$, and 
then 
breaking $U(1)$ via the vev $A^{12} \approx \epsilon' M$. The 
symmetry 
breaking 
\begin{equation}
 U(2) \stackrel{\epsilon}{\Rightarrow} 
U(1) \stackrel{\epsilon'}{\Rightarrow} 1
\label{eq:sb}
\end{equation}
produces the Yukawa coupling textures
\begin{equation}
M_{LR} = v \pmatrix{0 & \epsilon' & 0 \cr -\epsilon' & \epsilon &
  \epsilon \cr 0 & \epsilon & 1}.
\label{eq:MLR}
\end{equation}

\section{General Effective Theory of Neutrino Masses.}

Without right-handed neutrinos, the most general $U(2)$ effective
Lagrangian for neutrino masses, linear in $U(2)$ breaking fields, is
\begin{equation}
\mathcal{L}_{eff}^\nu = {1 \over M}l_3 l_3 hh + {1 \over M^2} 
\left(l_3 \phi^{a} l_a hh + l_a S^{ab}l_b hh \right).
\label{eq:nueff}
\end{equation}
where $l_a, l_3$ are lepton doublets.
The term $l_a A^{ab} l_bhh$ vanishes by symmetry; hence the above vevs
give the neutrino mass texture 
\begin{equation}
M_{LL} = {v^2 \over M} \pmatrix{0 & 0 & 0 \cr 0 & \epsilon &
  \epsilon \cr 0 & \epsilon & 1}.
\label{eq:MLLeff}
\end{equation}
so that the lightest neutrino is massless.\footnote{Including
operators higher order in the $U(2)$ breaking fields, 
the lightest neutrino remains
massless in a supersymmetric theory, but not in the non-supersymmetric
case , where operators such as $l_a A^{ab} \phi_b^\dagger l_3 hh$
occur.} 
The mixing angle for $\nu_\mu \rightarrow \nu_\tau$ oscillations,
$\theta_{\mu \tau}$,
is of order $\epsilon$ --- the same order as
mixing of the quarks of the two heavier generations, $V_{cb}$ --- and
is much too small to explain the atmospheric neutrino fluxes. However,
in theories with flavor symmetries, the seesaw mechanism typically
does not yield the most general neutrino mass matrix in the low energy
effective theory. This apparent problem requires that we look more
closely at the full theory, including the right-handed neutrinos.

\section{The Seesaw Mechanism: A Single Light $\nu_R$}




Adding three right-handed neutrinos to the theory, $N_a + N_3$, the
texture for the Majorana mass matrix is:
\begin{equation}
M_{RR} = M \pmatrix{0 & 0 & 0 \cr 0 & \epsilon &
  \epsilon \cr 0 & \epsilon & 1}.
\label{eq:MRR}
\end{equation}
with the 12 and 21 entries again vanishing by symmetry. In
supersymmetric theories the zero eigenvalue is not lifted at higher
order in the flavor symmetry breaking. This presents a problem for the
$3 \times 3$ seesaw mechanism in $U(2)$ theories, since
$M_{LL} = M_{LR} M_{RR}^{-1} M_{LR}^T$ and $M_{RR}$ cannot be 
inverted.

One approach \cite{CH} is to allow further flavor symmetry breaking
vevs, for example $\phi^1 \neq 0$, so that $M_{RR}$ has no zero
eigenvalues. Remarkably, taking $\phi^1/M \approx \epsilon'$, the 
seesaw 
gives
$\theta_{\mu \tau} \approx 1$, as needed for the atmospheric neutrino
anomaly. On the other hand, this pattern of neutrino masses cannot
explain the solar neutrino fluxes, and the additional flavor breaking
vevs remove two of the highly successful mass relation predictions of
the quark sector.

In this paper we keep the minimal $U(2)$ symmetry breaking vevs and
pursue the consequences of the light $N_e$ state which results from
(\ref{eq:MRR}). The singular nature of $M_{RR}^{-1}$ is not a problem;
it is an indication that $N_e$ cannot be integrated out of the
theory. However, $N_\tau$ and $N_\mu$ do acquire large masses, and 
when
they are integrated out of the theory the low energy $4 \times 4$
neutrino mass matrix is:
\begin{equation}
M^{(4)}=\pmatrix{ & & & 0 \cr & M_{LL} & & \epsilon' v \cr & & & 
0 \cr 0&
\epsilon' v & 0 & 0}
\label{eq:4b4general}
\end{equation}
where $M_{LL}$ is a $3 \times 3$ matrix in the $(\nu_{a},\nu_{3})$ 
space, determined from seesawing out the two heavy
right-handed states, and has one zero eigenvalue.

Because the $N_e - \nu_{\mu}$ mixing is weak scale, while all other
couplings to $\nu_\mu$ are suppressed, $N_e$ and $\nu_\mu$
are maximally mixed. Thus, 
we note that {\it a direct application of the $U(2)$ theory 
to
  the neutrino sector predicts a $45^o$ mixing between $\nu_\mu$
  and $\nu_s$!}

There is a significant phenomenological difficulty with 
this model. The mass of the $N_e - \nu_\mu$ pseudo-Dirac state is
of order $\epsilon' v$. Using a value for $\epsilon'$ extracted from 
an analysis
of the charged lepton sector, this is of order 1 GeV, well in
excess of the 170 keV limit obtained from direct searches.
One  simple solution is to restrict the couplings of the right-handed 
neutrinos by an additional $U(1)_N$ approximate flavor symmetry. Each
$N$ field carries $N$ charge +1, while the symmetry is broken by a
field with charge -1, leading to a small dimensionless breaking
parameter $\epsilon_N$. The entries in the neutrino mass matrices
receive further suppressions
\begin{eqnarray}
M_{LR} \Rightarrow \epsilon_N M_{LR} &
M_{RR} \Rightarrow \epsilon_N^2 M_{RR}
\label{eq:replacements}
\end{eqnarray}
which, for the $4 \times 4$ light neutrino matrix, simply leads to 
the replacement $\epsilon' v \Rightarrow \epsilon_N \epsilon' v$
in the $N_e - \nu_\mu$ entry, giving
\begin{equation}
  M^{(4)}=\pmatrix{{\epsilon'^2 \over \epsilon} {{v^2}\over
       M}  & {
     \epsilon'{{v^2}\over M}} & \epsilon' {{v^2}\over M}& 0
\cr \epsilon' {{v^2}\over M}
& \epsilon {{v^2}\over M}& \epsilon {{v^2}\over M} &
\epsilon_N \epsilon' v \cr
\epsilon' {{v^2}\over M} & 
\epsilon {{v^2}\over M}&  {{v^2}\over M} & 0 \cr 0 &  
  \epsilon_N \epsilon' v &0 & 0}
\label{eq:NMmass4by4}
\end{equation}
It is understood that all entries have unknown $O(1)$ coefficients.

Note that $M_{LL}$ is unchanged. There is a simple
reason for this. If we modify our right-handed couplings by the
replacements $M_{LR} \rightarrow M_{LR} T$, $M_{RR} \rightarrow T^T
M_{RR} T$, where $T$ is any diagonal matrix, then

\begin{equation}
M_{LL} \Rightarrow M_{LR} T (T^T M_{RR} T)^{-1} (M_{LR} T)^T = M_{LL}.
\label{eq: MLLinvariance}
\end{equation}
It is interesting that the observed value of $\delta m^2_\odot$ can give the
appearance that right-handed neutrinos receive GUT-scale masses,
while their masses are in fact much lower.

If the $N_e - \nu_\mu$ entry dominates the mass of
$\nu_\mu$, i.e. if $\epsilon_N \gg {v\over M}$, this $4 \times 4$
matrix splits approximately into two $2 \times 2$
matrices, and maximal mixing is preserved.
One $2 \times 2$ matrix describes the pseudo-Dirac state
\begin{equation}
\pmatrix{ \epsilon {v^2 \over M} &  \epsilon_N \epsilon' v \cr
  \epsilon_N \epsilon' v  & 0 }
\label{eq:NM2b2H}
\end{equation}
while $\nu_e \Rightarrow \nu_\tau$ mixing is described by
\begin{equation}
{v^2 \over M} \pmatrix{{{\epsilon'}^2 \over \epsilon} &
  \epsilon' \cr \epsilon' & 1}
\label{eq:NM2b2L}
\end{equation}
The resulting masses and mixings are given in Table \ref{tb:NMmm}.

\begin{table}[t]
 \renewcommand{\arraystretch}{1.5}
 \newcommand{\lw}[1]{\smash{\lower2.ex\hbox{#1}}}
 \begin{center}
  \begin{tabular}{|l|ll|ll|} \hline \hline
& $m_{light}$ & $m_{heavy}$  & $\delta m^2$ &
$\theta_{mix}$  \cr \hline
  (1) Heavy states & $v \epsilon_N \epsilon'  - \epsilon { v^2 \over 
2 M }$  
 & $v \epsilon_N \epsilon' +  \epsilon {v^2 \over 2 M }$ 
& ${v^3 \over  M} {\epsilon_N \epsilon \epsilon'}$ &
$45^{o}$ \cr
 
 (2) Light states & ${\epsilon'^2 \over \epsilon}{v^2 \over M}$ & 
${v^2
   \over M}$  & $({v^2 \over M})^2$ & $\epsilon' $

\cr 
\hline  \hline
\end{tabular}
\end{center}
\caption{General Theory: the masses, mixings, and splittings of the 
two sets of neutrinos.}
\label{tb:NMmm}
\end{table}

Since $\epsilon$ and $\epsilon'$ are determined by the charged fermion
masses, in the neutrino sector there are two free parameters, 
$\epsilon_N$ and
$M$, which describe five important observables: $\theta_\odot$,
$\theta_{atm}$, $\delta m^2_\odot$, $\delta m^2_{atm}$ and $m_\nu$, 
the
mass of the pseudo-Dirac muon neutrino. However, the various
predictions of the theory have varying levels of certainty. Because
there are a large number of order one constants in the original
formulation of the theory, we can end up with a prediction which has a 
coefficient of a product of some number of these quantities. To assess 
the level of certainty, we will include a quantity $i$, which we term
the ``stability index'' of the prediction, which is simply the power
of unknown order one coefficients appearing in the prediction.

Two of the three resulting predictions are the mixing angles
\begin{eqnarray}
\sin \theta_\odot \approx \epsilon' \hskip 0.25in \left[ i=4 \right], &  
\theta_{atm} = 45^{o} \hskip 0.25in \left[ i=0 \right].
\label{eq:NMangles}
\end{eqnarray}
The postdiction of a maximal mixing angle for atmospheric oscillations
is an important consequence of the $U(2)$ theory. 
The value of $\epsilon'$ extracted from the charged fermion sector is
$0.004$, within an order of magnitude of the central value
$\theta_\odot = 0.037$ of the recent BP98 fit to the solar data, and 
within a factor of $4$ of the minimal acceptable value of $0.016$ 
\cite{BP98}. 
Such a discrepancy is not a great concern, as we gain a comparable 
contribution from the
charged lepton matrix. Furthermore, the prediction of 
$\theta_{\odot}$ involves the fourth power of unknown order one 
coefficients, thus $i=4$, and is somewhat uncertain.   

The relevant mass splitting for the $\nu_e \rightarrow \nu_\tau$
oscillations occuring in the sun is
\begin{equation}
\delta m^2_\odot \approx \left({v^2 \over M} \right)^2.
\label{eq:m^2solar}
\end{equation}
While this is not a prediction of the theory, it is intriguing, as has
been noticed elsewhere in other contexts, that if $M$ is taken close
to the scale of coupling constant unification, $\delta m^2_\odot
\approx 10^{-5}$ eV$^2$, in the right range for either small or large angle MSW
oscillations. 

The final free parameter $\epsilon_N$ is fixed by the observed mass
splitting for atmospheric oscillations
\begin{equation}
\delta m^2_{atm} \approx \epsilon \epsilon' \epsilon_N {v^3 \over M}
     \approx  \epsilon \epsilon' \epsilon_N v \sqrt{\delta m^2_\odot} 
\label{eq:m^2atm}
\end{equation}
giving $\epsilon_N \approx 10^{-8}$ --- the $U(1)_N$ symmetry is
broken only very weakly.

The final prediction is for the mass of the heavy pseudo-Dirac
$\nu_\mu N_e$ state:
\begin{equation}
m_\nu \approx \epsilon' \epsilon_N v \approx 
{\delta m^2_{atm} \over \epsilon \sqrt{\delta m^2_\odot}}
\approx 10^{0.4} \mbox{eV} - 10^2 \mbox{eV}, \hskip 0.25in \left[ i=4 \right]
\label{eq:numass}
\end{equation}
where the given spread in mass is due to uncertainty in $\delta
m_{atm}^2$ and $\delta m_{\odot}^2$. While it is tempting to interpret 
this as a good candidate for hot dark matter, we will see later that
KARMEN places stringent limits on the acceptable values of $m_\nu$.

\section{A Variant Theory}

A variation on this breaking structure was explored in a particular 
model (see the second of \cite{u2GUTs}), and it is interesting to 
explore whether this same approach for neutrino masses can work 
within that model. In this variation, there is no $S^{ab}$ field 
present, and the $RR$ and $LR$ masses are given by

\begin{eqnarray}
M_{LR} = \pmatrix{ 0 & \epsilon' & 0 \cr -\epsilon' & 0 & \epsilon 
\cr 0 & \epsilon & 1} & M_{RR} = \pmatrix{ 0 & 0 & 0 \cr 0 & 0 & 
\epsilon \cr 0 & \epsilon & 1}
\label{eq:MsM}
\end{eqnarray}

\noindent generating a light $4\times 4$ mass matrix

\begin{equation}
  M^{(4)} = \pmatrix{
{\epsilon'^2 \over \epsilon^2} {{v^2}\over M}  
& {     \epsilon'{{v^2}\over M}} 
& {\epsilon' \over \epsilon} {{v^2}\over M}
& 0
\cr \epsilon' {{v^2}\over M}
& 0
& \epsilon {{v^2}\over M} 
& \epsilon_N \epsilon' v 
\cr {\epsilon' \over \epsilon} {{v^2}\over M} 
& \epsilon {{v^2}\over M}
&  {{v^2}\over M} & 0 
\cr 0 
&  \epsilon_N \epsilon' v 
& 0 
& 0}
\label{eq:Mmass4by4.1}
\end{equation}

This matrix is problematic, because the $2\times2$ submatrix for the
atmospheric neutrinos does not contain a splitting term. Of course, a
splitting would be generated through interactions with the other
left-handed states, we estimate 
\begin{equation}
M^{(4)}_{\mu \mu} \approx \epsilon^2 {1 \over m_\nu} ({v^2 \over M})^2.
\label{eq:Msplit.1}
\end{equation}
Consequently our atmospheric splitting is
\begin{equation}
\delta m^2_{atm} \approx \epsilon^2  ({v^2 \over M})^2.
\label{eq:Msplite.2}
\end{equation}

\noindent Since we have $ ({v^2 \over M})^2 = \delta m^2_\odot$, this would
predict $ \delta m^2_{atm} < \delta m^2_\odot$, which is unacceptable.
One simple solution is to allow the appearance of the 
operators 
\begin{eqnarray}
({1 \over M})^{2}¥ \phi^a \phi^b N_a N_b M_{GUT}¥ \\
({1\over M})^2  \phi^a \phi^b N_a \nu_b H.
\label{eq:newoperators}
\end{eqnarray}

The inclusion of one or both of these operators in our Lagrangian has
the same effect on our final mass matrix, inducing
$M^{(4)}_{\mu \mu} \approx \epsilon^2 {v^2 \over M}$ and yielding the $2 
\times 2$ submatrix

\begin{equation}
\pmatrix{ \epsilon^2 {v^2 \over M} &  \epsilon_N \epsilon' v \cr
  \epsilon_N \epsilon' v  & 0 }
\label{eq:M2b2H}
\end{equation}
describing the pseudo-Dirac state,
while $\nu_e \Rightarrow \nu_\tau$ mixing is now described by

\begin{equation}
{v^2 \over M} \pmatrix{{({\epsilon'\over \epsilon})^2 } &
  {\epsilon' \over \epsilon} \cr {\epsilon' \over \epsilon}  & 1}
\label{eq:M2b2L}
\end{equation}
The resulting masses and mixings are given in table \ref{tb:Mmm}.
\begin{table}[t]
 \renewcommand{\arraystretch}{1.5}
 \newcommand{\lw}[1]{\smash{\lower2.ex\hbox{#1}}}
 \begin{center}
  \begin{tabular}{|l|ll|ll|} \hline \hline
& $m_{light}$ & $m_{heavy}$  & $\delta m^2$ &
$\theta_{mix}$  \cr \hline
  (1) Heavy states & $v \epsilon_N \epsilon'  - \epsilon^2 { v^2 
\over 2 M }$  
 & $v \epsilon_N \epsilon' +  \epsilon^2 {v^2 \over 2 M }$ 
& ${v^3 \over  M} {\epsilon_N \epsilon^2 \epsilon'}$ &
$45^{o}$ \cr
 
 (2) Light states & $({\epsilon' \over \epsilon})^2 {v^2 \over M}$ & 
${v^2
   \over M}$  & $({v^2 \over M})^2$ & $\epsilon' \over \epsilon $

\cr 
\hline  \hline
\end{tabular}
\end{center}
\caption{Without S field: The masses, mixings, and splittings of the 
two sets of neutrinos.}
\label{tb:Mmm}
\end{table}

The mixing angles in this variation are predicted to be

\begin{eqnarray}
\sin (\theta_\odot) \approx {\epsilon'\over \epsilon} \hskip 0.25in
\left[ i=5 \right], & \theta_{atm} = 45^{o} \hskip 0.25in \left[ i=0 \right] 
\label{eq:Mangles}
\end{eqnarray}
As the pseudo-Dirac muon neutrino is still present, the atmospheric 
angle is unchanged. However, the solar angle is changed somewhat. 
We should note that values for $\epsilon$ and $\epsilon'$ extracted 
for a fit of this model are different than for those of the previous 
model. Using values from fits in the charged fermion sector, we have 
$\epsilon \approx 0.03$ and $\epsilon' \approx 5 \times 10^{-4}$ or
$\epsilon' \approx 2.4 \times 10^{-4}$ (depending on certain signs),
yielding $\theta_\odot \approx O(1.5 \times 10^{-2})$. Given
the number of $O(1)$ parameters involved, this is again quite 
consistent
with the BP98 small-angle MSW solution.

The solar splitting scale is unchanged, while the atmospheric 
splitting is further surpressed by a factor of $\epsilon$.

\begin{eqnarray}
\delta m^2_{atm} \approx \epsilon' \epsilon_N v \epsilon^2 
\sqrt{\delta m^2_\odot} 
\label{eq:Msplittings}
\end{eqnarray}
We fit this splitting again with the free parameter $\epsilon_N 
\approx 10^{-6} - 10^{-7}$. The resulting muon neutrino mass is then

\begin{equation}
m_\nu \approx {\delta m^2_{atm} \over \epsilon^2 \sqrt{\delta
    m^2_\odot}} \approx 10^{1.7} \mbox{eV} - 10^{3.5} \mbox{eV} \hskip 0.25in \left[ i = 5 \right]
\label{eq:Mnumass}
\end{equation}

Thus, while the explanations of the solar and atmospheric neutrinos 
remain, the neutrino becomes potentially dangerous in its 
cosmological implications. However, given the large
stability index of this prediction, there are large uncertainties in 
the prediction for its mass.

\section{KARMEN and LSND}

The presence of an additional sterile state makes it possible that a
signal would be seen in short baseline $\nu_\mu \rightarrow \nu_e$
oscillations, such as has been reported at LSND \cite{LSND}.
An estimate of the LSND mixing angle from the neutrino sector 
gives $\epsilon \epsilon' 
\delta m_{\odot}^{2}/\delta m_{atm}^{2}$, a 
very small result. Hence, this mixing originates 
from the charged lepton sector

\begin{equation}
        \theta_{LSND} = \sqrt{m_{e}\over m_{\mu}} \hskip 0.25in
        \left[ i= 0 \right].
\end{equation}
The precise predictions for $1-2$ mixing angles in the charged sector 
is an essential feature of the $U(2)$ flavor symmetry. In the quark 
sector it is highly successful. In the lepton sector, 
$\theta_{LSND} =   \sqrt{m_{e}\over m_{\mu}}$ is only useful if the 
neutrino mixing is either predicted or small, as in this theory. Recently,
the KARMEN experiment has placed limits on the allowed region for such 
oscillations, giving a limit $m_\nu \le 0.6 \mbox{ eV}$ \cite{KARMEN}. 
While the
prediction for $m_\nu$ has a large stability index in both the general 
theory as well as the variant theory, because the initial range for
$m_\nu$ is so high in the variant theory, it is disfavored by this bound.

The general theory is much safer, however. As we discuss in the
appendix, the uncertainty due to order one coefficients would allow it 
to satisfy the KARMEN bound. Such a result would likely coincide with 
higher values of $\delta m^2_\odot$ and lower values of
$\delta m^2_{atm}$.

\section{Astrophysical and Cosmological Implications}

%
%
%
%
%
%
%
%

There are three important cosmological implications of our theory.
\vskip 0.1in
\noindent {\bf 1.} We predict a small, but potentially significant 
amount of neutrino hot dark matter. The KARMEN bound limits us to a
$0.6 \mbox{ eV}$ neutrino, but because there are two massive states, it 
is still within the interesting region for HDM.

\noindent {\bf 2.} We predict abundances for light nuclei resulting 
from four light neutrino species. While newer data suggest 
$D/H$ ratios lie in the low end of the range previously thought, and 
thus $N_\nu < 4$, this is still an open question.

\begin{figure}[tbp]
  \begin{center}
    \leavevmode
    \psfig{file=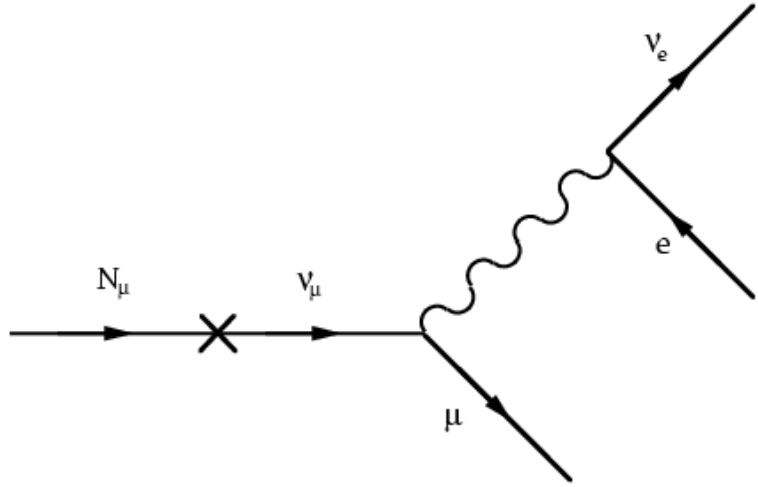,width=0.9\textwidth}
    \caption{Principal decay mode for $N_{\mu}$.}
    \label{fig:Ndecay}
  \end{center}
\end{figure}

\noindent {\bf 3.} There may be two further singlet neutrino states, 
dominantly $N_{\mu}$ and $N_{\tau}$, at or below the weak scale. 
Successful nucleosynthesis requires that they decay before the era of 
nucleosynthesis. Because the
mass eigenstates are slightly left-handed, the primary decay mode will 
be through the process shown in figure \ref{fig:Ndecay}. This is
similar to muon decay, which we use as a benchmark. For
the lighter of the two states, we estimate its lifetime to be

\begin{equation}
\tau_{N_{\mu}} \approx \epsilon'^{-2} \left( {\delta m_{atm}^{2} \over 
\delta m_{\odot}^{2}} \right)^{2} \left( {m_\mu \over m_{N_\mu}} \right)^5 
 \tau_\mu
\label{eq:lifetime}
\end{equation}
The mass of this particle is 

\begin{equation}
m_{N_\mu} \approx {1 \over \epsilon} {(\delta m^2_{atm})^2 \over
  (\delta m^2_{\odot})^{3/2}} \hskip 0.5in [i=12]
\label{eq:mNNM}
\end{equation}
for the general theory and

\begin{equation}
m_{N_\mu} \approx {1 \over \epsilon^2} {(\delta m^2_{atm})^2 \over
  (\delta m^2_{\odot})^{3/2}} \hskip 0.5in [i \approx 11-16]
\label{eq:mNM}
\end{equation}
in the variant theory. The stability index is approximate because it 
involves sums of order one coefficients of different powers. 
Furthermore, $i$ will change depending on which of 
(\ref{eq:newoperators}) are included.

The more dangerous case, the general theory,
then has a mass $O(100MeV)$ and thus a lifetime $\tau_{N_{\mu}}
\approx 10^3 s$, which is far too long to be acceptable. However,
because the lifetime has a fifth power dependence on the mass, and
because the prediction for the mass has index 12, deviations in the order one
quantities could very easily push the lifetime down to an acceptable
level. As we explore in the appendix, even conservatively we can only
reasonably estimate the mass of this particle to be in the range
$(17\mbox{MeV}, 40 \mbox{GeV})$, which means that the lifetime could easily be
$10^{-9} s$, without even beginning to push the limits of the order
one quantities. The details are presented in the appendix.

\section{Models}

The theory described in this paper has a low energy effective 
Lagrangian of (\ref{eq:basicu2}) for charged fermion masses, while the 
neutrino masses arise from the $U(2) \times U(1)_{N}$ effective 
Lagrangian

$$
{\mathcal{W}} = { \phi_{N} \over M}N_3 l_3 h + {\phi_N  \over M^2} \left( \
N_3 \phi^{a} l_a h + l_3 \phi^{a} N_a h+ N_a (S^{ab} + 
A^{ab}) l_b h \right)
$$
\begin{equation}
+
{\phi_N^2 \over M^3} \left( N_3 N_3  M + N_3 N_a \phi^a M + N_a N_b 
S^{ab} M \right)
\label{eq:nonrenormodel}
\end{equation}
where $N_{3}$ and $N_{a}$ have $U(1)_{N}$ charges $+1$, while 
$\phi_{N}$ has $U(1)_{N}$ charged $-1$.
The field $\phi_N$ gets a vev, breaking $U(1)_{N}$ 
and establishing an overall scale for these coefficients: 
${<\phi_{N}> \over M} = \epsilon_{N}$.
%
%
%
This effective theory can result from a renormalizable model by 
integrating out heavy states, both singlet and doublet under $U(2)$, 
in the Froggatt-Nielsen mechanism.

This symmetry structure on the right-handed singlet sector is far from 
unique. Another possibility is for $N_{a}$ to carry $U(1)_{N}$ charge, 
while $N_{3}$ is neutral under $U(1)_{N}$. This has no effect on any 
of our predictions, since the form of (\ref{eq:NMmass4by4}) for the 
light neutrino mass matrix is unchanged. The only change is that 
$N_{3}$ has a mass of the order of the unification scale $M$ rather 
than of order $\epsilon_{N}^{2}M$.

Another possible symmetry structure for the theory is 
$U(2)_{\psi}\times U(2)_{N}$, where $U(2)_{\psi}$ acts as usual on 
all the matter with non-trivial $SU(3)\times SU(2) \times U(1)$ 
quantum numbers, while $U(2)_{N}$ acts only on the three right-handed 
neutrinos, with $N_{3}$ a singlet and $N_{a}$ a doublet. The matrix 
$M_{RR}$ now has the form of (\ref{eq:MRR}), and arises from the 
renormalizable interactions

\begin{equation}
	W_{RR}= M N_{3} N_{3}+N_{3}\phi^{A} N_{A}+ N_{A}S^{AB}N_{B}
\label{eq:u2u2WRR}
\end{equation}
with vevs for $S^{22}$ and $\phi^{2}$ being of order $\epsilon M$ and 
breaking $U(2)_{N}\rightarrow U(1)_{N}$. The interactions for $M_{LR}$ 
are

\begin{equation}
	W_{LR}= l_{3}N_{3}h + {1 \over M} \left( l_{a} \phi^{a} N_{3} h +l_{3 }
	\phi^{A} N_{A}h+ l_{a}R^{aA}N_{A}h \right)
\label{eq:u2u2WLR}
\end{equation}
where $R^{aA}$ transforms as a (2,2). The vev for $R^{22}$ is also of 
order $\epsilon M$, since this is the scale of breaking of 
$U(2)_{\psi}\times U(2)_{N} \rightarrow U(1)_{\psi}\times U(1)_{N}$. 
The breaking scale for $U(1)_{\psi}$ is $\epsilon' M$, so the vev of 
$R^{12}$ takes this value. On the other hand, $U(1)_{N}$ is broken 
by $R^{21}$. We choose this scale to be smaller by a factor of 
$\epsilon_{N}$, $<R^{21}> \approx \epsilon_{N} \epsilon' M$, giving

\begin{equation}
	M_{LR}= v \pmatrix{0 & \epsilon' & 0 \cr \epsilon'  \epsilon_{N} &
	 \epsilon & \epsilon \cr 0 & \epsilon & 1}
	\label{eq:u2u2MLR}
\end{equation}
Integrating out the heavy states $N_{2}$ and $N_{3}$, which now have 
masses of order the unification scale, this theory now reproduces 
(\ref{eq:NMmass4by4}) for the mass matrix of the four light neutrinos.

The common features of these models, which are inherent to our scheme, 
are:

\begin{itemize}
	\item There is a $U(2)$ symmetry, which acts on the known matter 
	as $\psi_{3}\oplus \psi_{a}$, and is broken sequentially at scale 
	$\epsilon M$ and $\epsilon' M$.
	\item A $U(2)$ symmetry also acts on the three right-handed 
	neutrinos with $N_{3}$ a singlet and $N_{1,2}$ a doublet. This 
	$U(2)$, together with the symmetry of the Majorana mass, implies 
	that $N_{1}$ does not have a Majorana mass and becomes a fourth 
	light neutrino.
	\item There is an addition to the flavor group, beyond the $U(2)$ 
	which acts on $\psi$. At 
	least part of this additional flavor symmetry is broken at a scale 
	very much less than $M$, leading to a small Dirac mass coupling of 
	$\nu_{\mu} N_{e}$. Such a small symmetry breaking scale could be 
	generated by the logarithmic evolution of a scalar $m^{2}$ term.
\end{itemize}

\section {Conclusions}
%

There are several theories with sterile neutrinos \cite{Smirnov, 4x4s1, 4x4s2} 
some of which have $4 \times 4$ textures that split into two $ 2 
\times 2$ matrices. Such theories provide a simple picture for 
atmospheric oscillations via $\nu_{\mu}\rightarrow \nu_{s}$, and 
solar oscillations via $\nu_{e}\rightarrow \nu_{\tau}$, with $\delta 
m^{2}_{\odot} \approx {v^{2}\over M} \approx 10^{-5} \mbox{eV}^{2}$ 
for $M \approx M_{unif}$. However, theories of this kind typically do not 
provide an understanding for several key points:

\begin{itemize}
	\item Why is the Majorana mass of the singlet state $\nu_{s}$ small, 
	allowing $\nu_{s}$ in the low energy theory?
	\item Why does $\nu_{s}$ mix with $\nu_{\mu}$ rather than with 
	$\nu_{e}$ or $\nu_{\tau}$?
	\item Why is the $\nu_{s}-\nu_{\mu}$ state pseudo-Dirac, leading to 
	$45^{o}$ mixing?
	\item How can this extended neutrino sector be combined with the 
	pattern of charged quark and lepton masses in a complete theory of 
	flavor?
	\item What determines the large number of free parameters in the 
	neutrino sector?

\end{itemize}

In the theory presented here, all these questions are answered: the 
key tool is the $U(2)$ flavor symmetry, motivated several years ago by 
the charged fermion masses and the supersymmetric flavor problem. The 
simplest pattern of $U(2)$ symmetry breaking consistent with the 
charged fermion masses does not allow a Majorana mass for one of the 
three right-handed neutrinos. Furthermore, it is precisely this 
right-handed state that has a Dirac coupling to $\nu_{\mu}$ but not 
to $\nu_{e}$ or $\nu_{\tau}$, guaranteeing that $\nu_{\mu}$ is 
pseudo-Dirac with a $45^{o}$ mixing angle.

Our theory provides a unified description of both charged fermion and 
neutrino masses, in terms of just three small symmetry breaking 
parameters and a set of order unity coefficients. Some predictions, 
such as $|V_{ub}/V_{cb}| = \sqrt{m_{u}/m_{c}}$ and 
$\theta_{atm}=45^{o}$ are independent of the order unity coefficients 
and are precise. Other predictions, such as $|V_{cb}| \approx 
m_{s}/m_{b}$ and $\theta_{\odot} \approx \sqrt{m_{e} m_{\mu} / 
m_{\tau}^{2}}$ involve the order unity coefficients and are approximate. 
In the appendix we have introduced the ``stability index'' which 
attempts to quantify the uncertainty in such predictions according to 
the power of the unknown order unity coefficients appearing in the 
prediction.

There is one further free parameter of the theory---the overall mass 
scale $M$ setting the normalization of the right handed Majorana mass 
matrix. If $M$ is taken to be the scale of coupling constant 
unification $\delta m_{\odot}^{2} \approx 10^{-5}eV^{2}$.

The value of $\delta m_{atm}^{2}$ is not predicted--- 
this is the largest deficiency of the theory. It can be described by a 
very small flavor symmetry breaking parameter.
Once this parameter is set by the observed value of $\delta 
m_{atm}^{2}$, it can be used to predict the approximate mass range of the 
pseudo-Dirac $\nu_{\mu}$ to be in the range $10^{0.4} - 10^2
\mbox{eV}$, with significant additional uncertainty due to order one
coefficients. This, even with the KARMEN bound, allows for a neutrino
of cosmological interest with $\sum_{i} m_{\nu_{i}} \approx 1 \mbox{ 
eV}$. Such a neutrino could be seen at short
baseline experiments, and may have already been seen by LSND. 
Searching for $\nu_{\mu} \rightarrow \nu_{e}$, with $\sin ^{2}(2 
\theta ) = 2 \times 10^{-2}$, below the current limit of $\delta 
m^{2}$ is an important experiment for the $U(2)$ theory, since it is 
this prediction which differentiates $U(2)$ from several other 
theories with a light singlet neutrino.

\begin{table}[t]
 \renewcommand{\arraystretch}{1.5}
 \newcommand{\lw}[1]{\smash{\lower2.ex\hbox{#1}}}
 \begin{center}
  \begin{tabular}{|l|l|l|} \hline \hline

Experiment & Mode & Signal \cr
\hline

Present solar $\nu$ exp. & $\nu_e \rightarrow \nu_\tau$ & All data 
consistent with 2-flavor MSW
\cr
SNO & $\nu_e \rightarrow \nu_\tau$ & Confirm SK measurement of B8. 
Measure ${\phi_{NC} \over \phi_{CC}} \not = 1$
\cr
Borexino & $\nu_e \rightarrow \nu_\tau$ & Consistent with small-angle 
2-flavor MSW
\cr
KAMLAND & $\nu_e \rightarrow \nu_\tau$ & No signal
\cr
LSND,KARMEN& $\nu_\mu \rightarrow \nu_e$ & $\sin^{2} (2 \theta) = 2 
\times 10^{-2}$
\cr
K2K & $\nu_\mu \rightarrow N_e$ & $\nu_\mu$ disappearance. No 
$e$ appearance\cr
MINOS, ICARUS & $ \nu_\mu \rightarrow N_e$ & $\nu_\mu$ disappearance. 
No $\tau$ appearance. \cr
Atmospheric $\nu$ exp. & $\nu_{\mu} \rightarrow N_{e}$ & Confirm 2 
flavor $\nu_{\mu} \rightarrow \nu_{s}$ with $45^{o}$ mixing.

\cr

\hline  \hline
\end{tabular}
\end{center}
\caption{Experimental signals.}
\label{tb:signals}
\end{table}

Predictions of the theory for experiments sensitive to neutrino 
oscillations are listed in table \ref{tb:signals}. 
We expect a small angle MSW solution to the solar neutrino anomaly,
through a $\nu_e \Rightarrow \nu_\tau$ oscillation. The atmospheric
neutrino anomaly is from $\nu_\mu \Rightarrow \nu_s$. This will be
distinguishable from $\nu_\mu \Rightarrow \nu_\tau$ through a number
of means: LBL experiments will see $\nu_\mu$ disappearance, but no
$\nu_e$ or $\nu_\tau$
appearance. Improved statistics from Super-Kamiokande will be useful 
in 
distinguishing $\nu_\mu \Rightarrow \nu_\tau$ and $\nu_\mu 
\Rightarrow 
\nu_s$, for example via inclusive studies of multi-ring events
\cite{multiring}.

\appendix

\section{``Formalism" of the Stability Index}

It is difficult to establish a formalism for the stability index, 
because it involves an inherently ill-defined quantity, namely, what 
constitutes an order one quantity. However, the potential instability 
of various predictions to variations in these order one parameters 
makes some attempt to quantify this necessary. Such a quantification 
should be relatively insensitive to what precisely constitutes an 
``order one quantity".

Therefore, we demand the following quantities of the index:

\indent $\bullet$ An ``order one" quantity should be defined as a 
quantity $x$ with some probability distribution $P(x)$ to occur in an 
interval about 1. For reasons that will become clear later, it will 
be useful to consider instead the quantity ${\overline P}(y)$, where 
$x=10^y$.
\begin{itemize}
        \item This distribution should be ``sensible", namely 
        
        \indent 1.  $P(x)$ should be an even function in $Log(x)$, 
that is, ${\overline P}(y)$ is even in y. 

\indent 2.  ${\overline P}(y)$ should achieve its maximum 
value at 0.

\indent 3.  ${\overline P}(y)$ should have a spread 
characterized by its variance,
$$v^2 = \int_{-\infty}^{\infty} y^2 {\overline P}(y)$$
\indent the variance then quantifying what ``order one" is 
numerically.

\indent 4.  A product of two sensible distributions, 
correlated or uncorrelated, should be sensible. \vskip 0.15in
\item The index should have similar implications 
regardless of ${\overline P}(y)$, so long as it is sensible.
\item The definition of ${\overline P}(y)$ should be the 
only necessary input.

\end{itemize}

We shall explore the motivation for these assumptions and will 
shortly see that the presented index nearly meets the requirements, 
and with minor modifications can meet them enitrely.

We assume that the expectation value of $x$, and of any products of 
$x$, is unity. It follows immediately that ${\overline P}(y)$ should be 
even in $y$. We do not have strong arguments in favor of this 
assumption, and if it were relaxed, the formalism could be suitably 
modified.


For instance, consider the seemingly sensible distribution
$$ P(x) = \cases{{3 \over 8}, &if ${1 \over 3} \le x \le 3$;\cr 0, 
&otherwise. \cr }$$
\noindent which has been normalized to give total probability 1. The 
expectation value of a product of $n$ uncorrelated variables with 
such a distribution would be
\begin{equation}
<{\overline X}> = \int d^n x \prod_i P(x_i) x_i = ({10 \over 3})^n.
\label{eq:pathologicalP}
\end{equation}

\noindent Such a numerical pile-up of the central value of a product 
of order unity coefficients is excluded by our assumption.

What constitutes a ``sensible" distribution is, of course, a judgement 
call. Examples of what we consider sensible distributions would be 
\begin{itemize}
        \item Flat distributions taking on the value $1/a$ from 
$-a/2$ to $a/2$
\item Exponential distributions with standard deviation 
$\sigma$ 
\item Linearly decreasing distributions of the form 
$$ {\overline P}(y) = \cases{({1 \over a b}) (- {b \over a} |y| + b), 
&if  $-a \le y \le a$;\cr 0, &otherwise. \cr}$$
\end{itemize}
In fact, it can be shown that the last case it just the product of 
two uncorrelated quantities of the first type.

In all of these cases, the next moment $(x^4)$ is irrelevant in 
quantifying the likelihood of the variable being within a particular 
region about zero. Requirement 3 is then simply a statement that a 
sensible distribution should simply have one quantity, its variance, 
to determine how confident we are that the variable is within that 
region. This will then allow us to be more confident in deducing the 
significance of the variance of some product.

This being stated, we can actually go about constructing some 
approximation of confidence intervals. The ability to describe the 
distribution of one variable by its variance is useful in allowing us 
to calculate the variances for higher products. We begin by writing 
the formal expression for the probability distribution of $n$ 
uncorrelated variables $x_i = 10^{y_i}$ with probability 
distributions $\prod_i {\overline P}(y_i)$. We have
\begin{equation}
{\overline P}(z) = \int d^n y (\prod_i {\overline P}_i (y_i)) 
\delta(z-{\sum_i y_i})
\label{eq:pofz}
\end{equation}

This expression is tedious to calculate for given ${\overline P}(y)$, 
particularly for large $n$. However, its variance is a relatively 
simply calculation.
\begin{equation}
v_z^2 = \int dz {\overline P}(z) z^2 = \int dz \> d^ny (\sum_i y_i)^2 (\prod_i 
{\overline P}_i (y_i)) \delta(z-{\sum_i y_i})
\label{eq:zvariancestart}
\end{equation}

\noindent Expanding the squared term we find terms
\begin{equation}
\int dz\> d^ny \> y_i \> y_j (\prod_i {\overline P}_i (y_i)) 
\delta(z-{\sum_i y_i})= \cases{0, &for $i\ne j$; \cr v_i^2, &if 
$i=j$, \cr}
\label{eq:expandzvariance}
\end{equation}
giving

\begin{equation}
{v_z^2}_{uncorr} = \sum_i v_i^2
\label{eq:zvariance}
\end{equation}
For $n$ correlated variables, a similar calculations yields

\begin{equation}
{v_z^2}_{corr} = n^2 v_0^2
\label{eq:zvarrcorr}
\end{equation}
where $v_0^2$ is the variance of the original variable. 

Thus, a product of $n$ correlated order one quantities is far more 
unstable than a product of $n$ uncorrelated order one quantities. 
Simply counting the total number of order one coefficients is not 
sufficient. Thus we will refer to a product of the form
\begin{equation}
 \prod_i x_i^{n_i}
\label{eq: genproduct}
\end{equation}
as having index $(\sum_i n_i)$ of type $(n_1, n_2,..., n_m)$. If some
of the $n_i$ are repeated, we use the shorthand of writing $n^j$, if
$n$ is repeated $j$ times. We assume all order one quantities have the 
same distribution. A product of type $(n_1, n_2,..., n_m)$, has variance

\begin{equation}
v^2_{(n_i)} = v_0^2 \sum_i n_i^2
\label{eq:genvariance}
\end{equation}

This works extremely well for products of order one
coefficients. However, a sum of order one coefficients is not
necessarily order one. In these cases, it is usually best to perform a 
Monte Carlo to determine the distribution. 

\section{Sensible distributions}
To characterize the probability of a general product to be within a 
certain region about 1, it is necessary to explore the particular 
forms of various distributions. We consider three reasonable 
distributions to be i) the flat distribution, ii) the Gaussian 
distribution, and iii) the linearly decreasing distribution.

A product of two equal width flat distributions yields a linear 
distribution, so we need only consider the flat and Gaussian cases. 
Gaussian distributions are well understood: products of variables 
with Gaussian ${\overline P}(y)$ functions are again Gaussian, 
allowing standard statistical techniques to be applied.

Products of flat distributions very quickly become characterized by 
Gaussian distributions. We have performed explicit Monte Carlos for 
$n=1,2,3,5,7,9$ uncorrelated variables. Even by $n=2$ the 
Gaussian approximation is good, 
and for $n\ge 3$ it is very good. We thus believe it is reasonable to 
simply use Gaussian distributions, making a statistical 
interpretation of the variances simple.

For a standard, we propose using a distribution with variance $v = 
\sqrt{1 \over 12}$, which corresponds to the variance of a flat 
distribution for $ -{1 \over 2} \le y \le {1 \over 2}$. Changing the 
width of such a distribution from 1 to $a$ would amount to 
multiplying this variance by a. Such generally mild sensitivity of 
the index to variations in the initial distribution is one its 
desirable qualities. We can then take ``1-$v$" and 
``2-$v$" regions with $|y|\le v$ and $|y| \le 2v$, respectively. As should 
be clear, these should not be interpreted as the 
precise 67\% and 95\% 1 and 2-$\sigma$ regions, because $\sigma$ is 
not precisely defined. They are simply regions of medium and strong 
confidence, respectively.

As an example, consider a prediction with an unknown coefficient of order one 
quantities of the form $x_1^2 x_2 x_3$. We say this has index 
$2+1+1=4$ of type $(2,1,1)$, which we will write in shorthand as 
$(2,1^{2})$. Assuming the standard variance given 
above, this coefficient has variance $v = \sqrt{2^2 + 1^2 + 1^2 \over 
12} = \sqrt{1 \over 2}$. Thus, we can have medium confidence that the 
prediction for $x$ is known within a factor of $10^{v}=5$, and strong confidence 
the the prediction is within a factor of $10^{2v}=25$. 

We can also see that this reduces to the expected prediction in the 
case of a variable of index $1$ of type $(1)$. It will have variance 
$v = \sqrt{1\over 12}$ which gives medium confidence that the 
prediction is known within a factor of 1.9, and strong confidence it is 
known within a factor of 3.8. This is a good consistency check that 
the index predicts what we would expect in the case of a single order 
one coefficient.

\section{Reassessing the uncertainties in the $U(2)$ neutrino model}

In lieu of the preceding analysis, we address the index type of the
predictions already presented, and thus assess strong and medium
confidence regions of each prediciton. We list all uncertainties for 
the general theory in
table \ref{tb:NMuncertain}.

\begin{table}[t]
 \renewcommand{\arraystretch}{1.5}
 \newcommand{\lw}[1]{\smash{\lower2.ex\hbox{#1}}}
 \begin{center}
  \begin{tabular}{|l|ll|ll|} \hline \hline
 quantity & type & Sign & Range 
 of Med &
Range of Strong  \cr 
& (*=approx type) & convention & Confidence & Confidence \cr
\hline
  $\theta_{atm}$ & 0&  n/a & $\pi \over 4$ exact & exact \cr
$\theta_{\odot}$ & 4*& + & $(0.002,0.03)$& $(0.0006,0.1)$ \cr
$\theta_{\odot}$ & 4*& - & $(0.005,0.012)$& $(0.0001,0.06)$ \cr
$m_\nu$ & $(2,1^2)$ & none &$(2,50)\mbox{eV}$ & $(0.4, 250)\mbox{eV}$ \cr
$m_{N_L}$ &$(4,3,2^{2},1)$ & none & $(17\mbox{MeV},40\mbox{GeV})$ &$(300 
\mbox{keV},1.9 \mbox{TeV})$ \cr 
$m_{N_H}$ &$(4,2^{4})$ & none & $ (470 \mbox{MeV},870 
\mbox{GeV})$ &$ (11 \mbox{MeV},37 \mbox{TeV})$ \cr 

\hline  \hline
\end{tabular}
\end{center}
\caption{General Theory: uncertainties in predicitions. The regions
  listed here are simply for the uncertainty due to order one
  coefficients. Additional error due to uncertainty in input
  quantities, in particular in $m_\nu$, can also be significant.}
\label{tb:NMuncertain}
\end{table}

In the general theory with the S-field, the atmospheric mixing angle
is completely stable, while the solar angle is of approximate type 
$(1^4)=(1,1,1,1)$. However, it involves a sum of order one coefficients, 
motivating the use of Monte Carlos. Since a sum is involved, the 
relative sign of the order one quantities becomes relevant, and we 
list those cases seperately. These Monte Carlos allow us to claim 
that we have medium confidence that ${\theta_{\odot}}$lies 
within $(0.002, 0.03)$ and 
strong confidence that it lies within $(0.0001,0.1)$, giving large 
overlap of the BP98 region.

The mass of the pseudo-Dirac neutrino has stability index 5 of type $(2,1^3)$, 
giving a medium confidence to know this within a factor of 5, and strong
confidence within a factor of 25. Given the uncertainty in $\delta
m^2_{atm}$ and $\delta m^2_\odot$, which determine the prediction,
$m_\nu$ could conceivably be as low as $0.1 \mbox{eV}$.

The masses of the right-handed states are not known so well. The mass
prediction is, for the heavier state, of type $(4,2^4)$, and, for the
lighter state, of type $(4,3,2^2,1)$. This would give medium
confidence to know the masses at factors of 43 and 48,
and strong confidence at factors of 1800 and 2300, respectively. The
cosmological implications of these neutrinos are very uncertain, given 
that the lighter could be well over a TeV in mass.

Without the S field, certain uncertainties change. The precise nature 
of the changes depends on which splitting operators are included and 
what sign convention is taken. Because of the large number of 
permutations, we list only the basic results. The atmospheric angle 
is, as expected, completely certain. The solar angle becomes slightly 
more uncertain, but still overlaps BP98 well. The heaviest two 
righthanded masses typically become less certain by a factor of 
roughly 100, but the uncertainty is so large that the 
phenomenological predictions remain the same. The only dramatic 
difference in the 
variant theory is that $\nu_{\mu}$ has a medium confidence region on 
its mass of $(24 \mbox{eV},4 \mbox{keV})$,  and a strong confidence region of 
$(2 \mbox{eV},48 \mbox{keV})$. Including the uncertainties in the
input quantities, the mass could be as low as $0.4 eV$, which 
escapes the KARMEN bound, although narrowly.

\section*{Acknowledgements} 

This work was supported in part by the
U.S. Department of Energy under Contracts DE-AC03-76SF00098, in part
by the National Science Foundation under grant PHY-95-14797.

\end{document}